\documentclass[12pt]{article}

\usepackage{latexsym,amsmath,amssymb,graphicx}
\usepackage{graphicx}

\evensidemargin \oddsidemargin

\begin{document}


\begin{titlepage}

\renewcommand{\thefootnote}{\fnsymbol{footnote}}



\vspace{15mm}
\baselineskip 9mm
\begin{center}
  {\Large \bf Semiclassical approach for the evaporating black hole revisited}
\end{center}

\baselineskip 6mm
\vspace{10mm}
\begin{center}
  Yongwan Gim${}$\footnote{Email: yongwan89@sogang.ac.kr} and  
  Wontae Kim${}$\footnote{Email: wtkim@sogang.ac.kr}
  \\
  \vspace{3mm}
  {\sl ${}$Department of Physics, Sogang University, Seoul 121-742,  Republic of Korea}

\end{center}

\thispagestyle{empty}

\vfill
\begin{center}
{\bf Abstract}
\end{center}
\noindent
A recent calculation shows that the observed energy density in the Unruh
state at the future event horizon as seen by a freely falling
observer is finite if the observer is released from rest at any
positive distance outside the horizon; however,
it is getting
larger and larger so that
it is negatively divergent at the horizon in the limit that the
observer starts falling from rest at the horizon,
which corresponds to the infinite boost with respect to the freely falling observer
at a finite distance from the horizon.
In order to resolve some conflicts between the recent calculation and the conventional ones in the well-known literatures,
the calculation of the free-fall energy density is revisited and
some differences are pointed out.\\ [5mm]
\vspace*{1cm}

\vspace{20mm}

\vfill
\end{titlepage}

\baselineskip 6.6mm
\renewcommand{\thefootnote}{\arabic{footnote}}
\setcounter{footnote}{0}

\section{introduction}
\label{introduction}
One of the most outstanding works over the last four decades
in the quantum theory of gravity is the derivation of
Hawking radiation from black holes \cite{Hawking:1974sw}
since it has provided many profound questions and intriguing puzzles about
the quantum theory of gravity
such as
information loss problem \cite{Hawking:1976ra}, black hole complementarity
\cite{Susskind:1993if,  Stephens:1993an, Susskind:1993mu}, and
the recent firewall paradox \cite{Almheiri:2012rt,Almheiri:2013hfa}
which states
that a freely falling observer when crossing the horizon encounters firewalls
which are high frequency outgoing quanta near the horizon and the infalling observer
burns up \cite{Almheiri:2012rt,Almheiri:2013hfa}.
Subsequently, various aspects related to the firewall issue have
been extensively studied in Refs.\cite{Mathur:2009hf, Bousso:2012as,Nomura:2012sw,Page:2012zc,Giddings:2012gc,
JACOBSON:2013ewa,Kim:2013fv,Maldacena:2013xja, Hotta:2013clt}.
The presence of the firewalls has something to do with the failure
of the equivalence principle or
breakdown of semiclassical physics at macroscopic distance from the horizon,
which eventually makes black hole complementarity incomplete.
A similar prediction referred to as an energetic
curtain has also been done based on different
assumptions~\cite{Braunstein:2013bra}.
On the other hand, it has been claimed that
firewalls are not essential since the unitary evolution of black hole
entangles a late mode located outside the event horizon with a combination
of early radiation and black hole states at the same time \cite{Hutchinson:2013kka},
and also argued that the remaining set of nonsingular realistic states do not have
firewalls but yet preserve information \cite{Page:2013mqa}.
Of course, it should be emphasized that a static outside observer would not 
witness formation of the horizon and always sees unitary evolution \cite{Saini:2015dea} without
any firewall-like object. 
Since all these issues are related to the freely falling observer,
the revisit of the free-fall energy density becomes so important.
Note that
the classic work \cite{Candelas:1980zt} tells us that the energy density in the Unruh state \cite{Unruh:1980cg} and
the Israel-Hartle-Hawking state \cite{Israel:1976ur,Hartle:1976tp}
are finite on the future horizon,
although it is divergent in the Boulware state  \cite{Boulware:1404}
on the future horizon.
Especially, for the black hole in the Unruh state,
the energy-momentum tensors were calculated
at the bifurcation two-sphere
in virtue of the vanishing affine connections
since the energy-momentum tensors could be regarded as quantities in the locally
flat spacetimes \cite{Candelas:1980zt}.
This calculation was in turn extended to the future horizon
by taking into account of a symmetry argument for the infinite time,
 and the finite energy density was eventually obtained
on the future horizon when the observer is dropped at the future event horizon {\it without any journey}.
The geodesic solution was not used in this work
so that the conclusion seems to be dubious
if the employed coordinates are not the local free-fall coordinates.
On the other hand, 
quantum gravitational collapse was investigated from the point of view of an 
infalling observer in order to investigate the quantum-mechanical modification of the 
collapse and the singularity \cite{Greenwood:2008ht}. 
And it was also pointed out that the observers dropped from
a finite distance outside the horizon will detect a finite amount of
radiation when crossing the horizon
by using the effective temperature method \cite{Barbado:2011dx, Barbado:2012pt}.
Recently, the free-fall energy density in
the soluble two-dimensional Schwarzschild black hole \cite{Eune:2014eka}
was directly calculated at arbitrary free-fall positions
in order to study the dependence of initial free-fall positions
of the free-fall energy density on the horizon and clarify
whether the freely falling observer could encounter
something non-trivial effect at the horizon or not. In particular,
solving the geodesic equation of motion exactly
over the whole region outside the horizon in the Unruh state \cite{Unruh:1980cg},
it could be shown that there exists the negative energy density up to
the extent to the exterior to the horizon of the black hole, roughly $r \sim 3M$
\cite{Eune:2014eka}, where the similar argument in connection with
the firewall argument appears in Ref. \cite{Freivogel:2014dca} and the
behavior of the energy density was also explicitly shown in Ref. \cite{Singh:2014paa}.
Note that the negative energy density is getting larger and larger
when the initial infalling position from being at rest approaches the horizon
\cite{Eune:2014eka}.
So, the energy density in the freely falling frame
is finite unless the observer is not dropped at the horizon.
However, if the observer were dropped at the horizon, the energy-density
 would be divergent. So,
  one might wonder why the behavior of the energy density in the freely falling frame at the horizon in Refs.
\cite{Eune:2014eka} is different from the result in Ref. \cite{Candelas:1980zt}.

In this work, we would like to elaborate what the differences
between the finite energy density calculated on the future horizon \cite{Candelas:1980zt} and
the recent calculations for the divergent energy density on the future horizon \cite{Eune:2014eka}.
Usually,
the Kruskal coordinates might be used in calculating the free-fall energy density
on the horizons; however,
in Sec. \ref{Kruskal coordinates},
we shall prove that affine connections derived in the
Kruskal coordinates do not always vanish at
$r=2M$, specifically
along the ingoing direction on the future horizon.
It means that it is impossible to regard the energy density calculated
in the Kruskal coordinates
as the one observed in the
freely falling frame, since the Kruskal coordinates
do not play a role of the local inertial coordinates any more,
so that the Kruskal time is no longer proper time on the future horizon.
Thus we should consider the general coordinate transformation of the energy-momentum tensors
from the fixed coordinates such as the Kruskal coordinates
to the local inertial coordinates, which gives
 the appropriate definition
for infalling energy density observed in the local inertial frame \cite{Ford:1993bw}
in Sec. \ref{free-fall energy density}.
After all, we will show that
the observed energy density in the Unruh
state at the future event horizon as seen from a freely infalling
observer is finite if the observer is released from rest at any
positive distance outside the horizon, where the result is compatible with
the one in Refs. \cite{Barbado:2011dx, Barbado:2012pt}; however, it is only
divergent at the horizon in the limit that the freely falling
observer starts falling from rest at the horizon \cite{Eune:2014eka}.
In Sec. \ref{blue shift}, the origin of the divergence
will also be mentioned from the viewpoint of the infinite blue shift
which is related to the infinite boost with respect to relatively
infalling observer at
a finite distance from the horizon.
Finally, the conclusion and comment
will be presented
in Sec. \ref{conclusions}.

\section{Kruskal coordinates}
\label{Kruskal coordinates}

To explain the reason why the different behavior of the energy density appears between
the recent calculation \cite{Eune:2014eka} and the classic work
\cite{Candelas:1980zt} with the well-known text book \cite{Birrell:1982ix},
we would like to present a heuristic calculation in terms of
the Callan-Giddings-Harvey-Strominger model \cite{Callan:1992rs},
where the length element is given as $ds^2=-e^{2\rho}dx^+dx^-$
together with the metric component of
$e^{-2\rho}=M/\lambda-\lambda^2 x^+ x^-$ in the Kruskal coordinates.
The Kruskal coordinates are
related to the tortoise coordinates through the coordinate transformations of
$2\lambda t=\ln(-x^+/x^-)$ and  $2\lambda r^*=\ln(-\lambda^2 x^+ x^-) $,
where $r^*=r+   (1/2\lambda) \ln[1-(M/\lambda) e^{-2\lambda r}]$.
The affine connections in the Kruskal coordinates are straightforwardly calculated as
$\Gamma^+_{++} (x^+,x^-) =2 \partial_+ \rho(x^+,x^-) \sim x^-$,
$\Gamma^-_{--} (x^+,x^-)=2 \partial_- \rho(x^+,x^-) \sim x^+$.
Note that the affine connection of $\Gamma^-_{--} (x^+,0)$ on the future horizon of
$x^-=0$ does not vanish, while
 $\Gamma^+_{++} (x^+,0) =0$. So, the geodesic equation of motion
tells us that $x^-$ cannot be a local flat coordinate on the future horizon.
However, the affine connections vanish at $x^{\pm}=0$
corresponding to the bifurcation point.

The awkward situation is not restricted to the above case, and it also
happens in the other models such as the two-dimensional Schwarzschild black hole
 which is actually
of our concern since the model is simple but it  shares most properties in
realistic four-dimensional black holes. The length element
is given as $ds^2=-f(r)dt^2  + f^{-1}(r)dr^2 $ with the metric function of $f(r)=1-2M/r$
in the Schwarzschild coordinates.
The conformal factor for the length element of $ds^2=-e^{2\rho}dx^+dx^-$ in the Kruskal coordinates is obtained as
\begin{equation}
\label{metric2}
e^{2\rho(x^+,x^-)}=\frac{2M}{r}e^{1-\frac{r}{2M}},
\end{equation}
from the conformal transformation of
$x^\pm=\pm 4Me^{\pm \sigma^\pm/4M}$,
where $\sigma^\pm=t\pm r^*$ and
$r^*=r-2M+2M\ln(r/2M-1)$.
The corresponding coordinate transformations
are implemented by
 $t=2M \ln(-x^+/x^-)$ and $r^*=2M\ln(-x^+x^-/(16M^2))$.
At first glance, the affine connections calculated from Eq. \eqref{metric2}  might be expected to vanish
at $r=2M$ since it is
given as
\begin{equation}
\label{ambiguity}
\Gamma^\pm_{\pm\pm} (t,r) =\mp \left(\frac{1}{2r}+\frac{M}{r^2}\right)
\sqrt{\frac{r}{2M}-1}
~~e^{\frac{\mp t-r+2M}{ 4M}}
\end{equation}
in the Kruskal coordinates. But this is not the case except the finite $t$ since
the vanishing square root and the divergent exponential function in $\Gamma^-_{--}$
compete on the future horizon.
So, it might be tempted that the affine connections at $t \rightarrow \infty$
would vanish on the future horizon away from the bifurcation two-sphere.
However, this is not the case as seen from Eqs. \eqref{limGppp} and \eqref{limGmmm},
and hence the two limits such as $r=2M$ and $t \rightarrow \infty$
should be taken at one stroke in order to justify the flatness via
affine connections on the future horizon.
For this purpose, if we take advantage of
the light cone expressions in the Kruskal coordinates,
then the affine connections \eqref{ambiguity} are neatly calculated as
\begin{align}
\Gamma^+_{++} (x^+,x^-) &= \frac{1}{x^+}\left(\frac{1}{\left(1+W(Z)\right)^2}-1\right), \label{Gppp2} \\
\Gamma^-_{--} (x^+,x^-) &=\frac{1}{x^-}\left(\frac{1}{\left(1+W(Z)\right)^2}-1\right)  \label{Gmmm2},
\end{align}
in virtue of the Lambert $W$ function defined as $Z=W(Z) e^{W(Z)}$ where $Z=-x^+x^-/(16M^2)$.
As a result, the affine connections on the future horizon of $x^-=0$ are written as
\begin{align}
\lim_{x^-\rightarrow0}\Gamma^+_{++} (x^+,x^-)  &=0, \label{limGppp}\\
\lim_{x^-\rightarrow0}\Gamma^-_{--}  (x^+,x^-) &= \frac{x^+}{8M^2} \ne 0, \label{limGmmm}
\end{align}
where  we used the relation of
$W(Z)=Z -Z^2 - O(Z^3)$ near the future horizon.
Note that $\Gamma^-_{--}$
does not vanish on the future horizon, and it turns out that the coordinate $x^-$ cannot be a
freely falling coordinate.

As expected,  these two-dimensional results
can also be applied to the four-dimensional Schwarzschild metric whose length element
is given as
$ds^2=-e^{2\rho}dx^+dx^- + r^2 ( d \theta^2 +\sin^2 \theta d \phi^2) $, where
$\rho$ and $r$ are functions of $x^{\pm}$.
The corresponding nonvanishing affine connections on the future horizon are illustrated such as
$\Gamma^-_{--}=x^+/(8M^2),
~\Gamma^+_{\theta\theta}=-x^+/2,
~\Gamma^+_{\phi\phi}=(-1/2)x^+ \sin^2\theta,
~\Gamma^\theta_{-\theta}=-x^+/(16M^2),
~\Gamma^\theta_{\phi\phi}=-\cos\theta\sin\theta,
~\Gamma^\phi_{-\phi}=-x^+/(16M^2)$, and
$\Gamma^\phi_{\theta\phi}=\cot\theta$.
We can choose $\theta = \pi/2$ since we are concerned with
the freely falling motion which is confined on the plane, but there still exist nonvanishing
affine connections on the future horizon.
In the light of these calculations, the Kruskal coordinates could not be local flat coordinates on the future
horizon except the bifurcation point joining the past horizon and the future horizon corresponding to
$x^{\pm}=0$.
Note that in Ref. \cite{Candelas:1980zt},
the energy-momentum tensors were calculated on the bifurcation two-sphere
for which
$\Gamma^\pm_{\pm\pm} (t,2M)=0$
for any finite time,
and in turn extended the analysis to the future horizon by taking infinite time
with a symmetry argument without any coordinate transformation to the local inertial frames.
Thus, if the energy-momentum tensors were calculated in the Kruskal coordinates
on the future horizon,
they could not be identified with the energy momentum tensors
in the freely falling frame at that point. So the finiteness of the
energy momentum tensors at the future event horizon should be reexamined.
\section{energy density in freely falling frame}
\label{free-fall energy density}

At the asymptotic infinity,
the energy-momentum tensors are easily defined in virtue of the tortoise coordinates since
they are more amenable to impose some boundary conditions compared to the other coordinate systems.
Let us now consider the energy-momentum
tensors in the tortoise coordinates, and
assume that the tensor transformations can be well-defined
from the tortoise coordinates to the Kruskal coordinates as a true tensors without
any anomalies,
\begin{align}
T_{\pm\pm}(x^+,x^-) &=\left(\frac{\partial \sigma^\pm}{\partial x^\pm} \right)^2 T_{\pm\pm}(\sigma^+,\sigma^-),\label{Tpp}\\
T_{+-}(x^+,x^-) &=\left(\frac{\partial \sigma^+}{\partial x^+} \right)\left(\frac{\partial \sigma^-}{\partial x^-} \right) T_{+-}(\sigma^+,\sigma^-).\label{Tpm}
\end{align}
If the energy-momentum tensors were calculated in the Kruskal coordinates,
they could not be identified with those observed by the freely falling observer
except the bifurcation point as discussed in the previous section.
Thus the coordinates $x^{\pm}$ should be replaced by the
local inertial coordinates in such a way that  the energy density in the local inertial coordinate
should be written as
 \begin{equation}
 \label{thanks}
 \epsilon=T_{\tau \tau}=\frac{d\sigma^{\mu}}{d\tau}  \frac{d\sigma^{\nu}}{d\tau}   T_{\mu\nu},
 \end{equation}
where $\tau$ is a proper time \cite{Ford:1993bw}.
In other words, the energy-momentum tensors \eqref{Tpp}
and \eqref{Tpm} calculated in the
Kruskal coordinates should be transformed to the local inertial coordinates.
Note that such a form of the energy density \eqref{thanks}
was already introduced  in order to exhibit the finite infalling
energy density on the future horizon in
Ref. \cite{Birrell:1982ix}.
The authors considered an observer moving along a line of constant Kruskal position of $x^1=a$ where
$a$ is a constant along with the two-velocity of $(u^{0},~ u^{1})=(dx^0 /d\tau,~dx^1/d\tau)= e^{-\rho}(1,0)$.
The constant spacial radius was expressed in the light cone coordinates as
$ x^+ =x^- +2a$ in the Kruskal coordinates.
However, the constant line does not obey
the geodesic
equation of motion but it can be a geodesic
solution only at the bifurcation point for which $a=0$.
Thus the calculation does not warrant the finiteness of the energy density
in the freely falling frame at the future horizon even in spite of
the correct definition of the infalling energy density \eqref{thanks}
\footnote{We have repeatedly been asked why our result in this work is incompatible with the result of Ref. \cite{Birrell:1982ix}
in the well-known text book. Our answer is: ``The proposed geodesic curve  in section 8.2 does not actually satisfy the
geodesic equation of motion, so that the energy density by the freely falling
observer discussed in Ref. \cite{Birrell:1982ix} is not the
free-fall energy density".}.

Now, it becomes clear why we have to use
the above definition of the infalling energy density along with
the correct geodesic
solution in order to study the energy density
in the freely falling frame.
Using Eq. \eqref{thanks},
we are going to
calculate the infalling energy density in the two-dimensional Schwarzschild black hole
in the Unruh state by means of the light-cone coordinates
in order to avoid any ambiguities in connection with
the future horizon.
Let us now start with the conformal gauge fixed energy-momentum tensors \cite{Christensen:1977jc},
\begin{align}
T_{\pm\pm} &=-\kappa [(\partial_\pm \rho)^2-\partial_\pm^2 \rho+t_\pm], \label{Tpp1}\\
T_{+-} &= -\kappa \partial_+ \partial_- \rho \label{Tpm1},
\end{align}
where they can be derived from the covariant conservation law and the
two-dimensional trace anomaly
for the number  of $N$ massless scalar fields
and $t_{\pm}$ are the integration functions and $\kappa=N/12 $.
The conformal factor of the two-dimensional Schwarzschild black hole
from Eq. \eqref{metric2}
is written as
\begin{equation}\label{metrictor}
e^{2\rho(\sigma^+,\sigma^-)}=1-\frac{2M}{r(\sigma^+,\sigma^-)},
\end{equation}
in terms of the tortoise coordinates,
 where the radial coordinate is also expressed as
$r(\sigma^+,\sigma^-)=2M(1+W(Y))$ and
by definition $Y=\exp[(\sigma^+-\sigma^-)/4M]$.
From Eqs. \eqref{Tpp1}, \eqref{Tpm1} and \eqref{metrictor},
it is easy to obtain the energy momentum tensors,
\begin{align}
T_{++} &=-\frac{\kappa}{64M^2}\frac{1+4W(Y)}{(1+W(Y))^4}, \label{Tpp2} \\
T_{--} &=-\frac{\kappa}{64M^2}\left(\frac{1+4W(Y)}{(1+W(Y))^4}-1\right), \label{Tmm2}\\
T_{+-} &=\frac{\kappa}{16M^2}\frac{W(Y)}{(1+W(Y))^4}  \label{Tpm2},
\end{align}
which satisfy the Unruh state because we chose $t_+=0$ and $t_-=-1/(64M^2) $
\cite{Unruh:1980cg}.
So, the ingoing flux is negative finite on the past horizon from Eq. \eqref{Tpp2}
and hence it is singular in the Kruskal coordinates, while
there is no outgoing flux on the future horizon from Eq. \eqref{Tmm2} so that
it is finite in the Kruskal coordinate on the future horizon.

Next, the components of the two-velocity
are obtained
by exactly solving the geodesic equation of motion for a massive particle as
\begin{equation}\label{upm}
u^\pm (\sigma^+, \sigma^- ; \sigma_s^+, \sigma_s^- )
=\left(\sqrt{1-\frac{1}{1+W(Y_s)}}\pm \sqrt{\frac{1}{1+W(Y)}-\frac{1}{1+W(Y_s)}}\right)^{-1},
\end{equation}
where the initial infalling position at
rest is denoted by $\sigma^{\pm}_s$, and $Y_s=\exp[(\sigma^+_s-\sigma_s^-)/4M]$.
From Eqs. \eqref{Tpp2},\eqref{Tmm2}, \eqref{Tpm2}, and \eqref{upm}, the energy density \eqref{thanks}
measured on the future horizon is given as
\begin{equation}\label{density}
\epsilon(\sigma^+, \sigma^- \to \infty ; \sigma_s^+, \sigma_s^- )
 = -\frac{\kappa}{256M^2 W(Y_s)}-\frac{33\kappa}{256M^2}+O(W(Y_s)),
\end{equation}
where the initial infalling position is assumed to be near the future horizon.
It is interesting to note that it is independent of $\sigma^{+}$, and
just depends on the initial infalling position $\sigma^{\pm}_s$.
From Eq. \eqref{density}, it turns out that there is no divergence unless we require
that the observer be at rest at the horizon.
On the other hand,
if the observer is dropped extremely on the future horizon for which
$Y_s$ and $W(Y_s)$ vanish, then the energy density
is negative divergent.

\section{blue shift}
\label{blue shift}
So far we have calculated the energy density in the freely
falling frame in the Unruh state
near the future horizon.
Let us now discuss, in particular, the
origin of the divergence when the observer is dropped
at the horizon as an extreme limit.
Considering a freely falling observer at the initial infalling position of $r_s$
without any journey for simplicity,
the energy density \eqref{thanks}
is written as
$\epsilon(r_s; r_s) =T_{tt}u^t u^t +T_{rr}u^ru^r  +2T_{tr}u^t u^r$
in the Schwarzschild coordinates.
When the infalling happens at rest i.e.,  $u^r|_{r_s}=0$,
then the infalling energy density at that moment
is reduced to $\epsilon(r_s;r_s) =(1/f(r_s))T_{tt}$
in virtue of $u^{t}|_{r_s}=dt/ d\tau|_{r_s} =1/\sqrt{f(r_s)}$.
Note that the red-shift factor is also responsible for
the gravitational time dilation which is larger and larger
as the initial infalling position approaches the horizon.
Next, the value of $T_{tt}$ in the Schwarzschild coordinates can be
directly obtained
by the use of  the coordinate transformation from
 the tortoise coordinates
to the Schwarzschild coordinates, then the energy density \eqref{thanks} becomes
\begin{equation}
\label{red}
\epsilon(r_s;r_s) =\left.\frac{1}{f(r_s)}[T_{++}+T_{--} +2 T_{+-}] \right|_{r_s},
\end{equation}
where the last term is independent of the vacuum state of black hole and
it can be written as $T_{+-} \sim -(\kappa/(16M^2))f(r_s)$ near the horizon.

When the initial infalling position extremely approaches the horizon $r_s \to r_H$,
Eq. \eqref{Tpp1}
can also be expanded asymptotically for each of vacuum states.
 First,
the leading order of contributions to the energy-momentum tensors
 in the Boulware state described by choosing $t_{\pm}=0$  \cite{Boulware:1404}
becomes finite since $T^{\rm{B}}_{\pm \pm} \sim -\kappa/(64M^2)$,
so that the energy density
\eqref{red} is divergent at the horizon.
For the Israel-Hartle-Hawking state implemented by choosing $t_+=t_-=-1/(64M^2) $  \cite{Israel:1976ur,Hartle:1976tp},
the leading order of energy-momentum tensors is written as
$T^{\rm{H}}_{\pm\pm} \sim -(\kappa/(16M^2))f(r_s)$ which
vanish asymptotically at the horizon;
however, the energy density is finite due to the redshift factor
in the denominator in Eq. \eqref{red}.
Hence, these two states result in drastically different conclusions.
By the way, in the Unruh state characterized by $t_+ =0$ and $t_-= -1/(64M^2)$,
the leading order of the energy-momentum tensors near the horizon is calculated asymmetrically as
 $T^{\rm{U}}_{++} \sim -\kappa/(64M^2)$ and
$T^{\rm{U}}_{--} \sim -(\kappa/(16M^2))f(r_s)$,
where the ingoing flux is negative finite while
the outgoing one vanishes at the horizon.
However, the energy density observed in the freely falling frame
at the horizon is divergent because
the negative finite ingoing flux $T^{\rm{U}}_{++}$ is
infinitely blue shifted just like the case of the Boulware state.
Actually, in this case,
the infinite boost is required with respect to the freely falling observer from a finite distance. So the divergent effect is from
moving at the speed of light relative to any infalling
frame that comes from any positive distance outside the horizon.
Thus the divergence is easily explained as a blueshift effect from moving at
the speed of light through radiation.

\section{conclusions}
\label{conclusions}
We showed that the Kruskal coordinates could not
be local inertial coordinates on the future horizon
by invoking nonvanishing affine connections.
So we investigated the energy density observed by the freely falling
observer in the Unruh state
by means of the proper definition
of the energy density.
The energy density observed at the future horizon
by the freely falling observer from rest is finite unless the observer is dropped
at the horizon. For the extreme case of the observer dropped at the horizon,
the energy density is divergent, which is due to the infinite blue shift of
the energy density.
The closer the initial infalling position approaches the horizon, the
more negative energy density appears.
Our calculations show that the energy density observed by the
freely falling observer at the future horizon is sensitive to both the
vacuum state of black hole and the initial infalling position of
the freely falling observer.
In Ref. \cite{Candelas:1980zt}, the author concluded that the energy density
in the fixed coordinate at the horizon
is finite without specifying the geodesic solution and initial free-fall position.
And in Ref. \cite{Birrell:1982ix} the correct definition of the energy density
was used; however, the geodesic curve
does not satisfy the geodesic equation of motion.
Therefore, the correct geodesic trajectory and the definition of the free-fall energy
density were used in this work compared to the well-known literatures.

The final comment is in order.
It is expected to observe
some amount of the energy density near the horizon,
which amounts to the curvature scale of
$ \sim 1/M^2$.
Thus the energy density in the freely falling frame
can be written as the improved form by the red shift factor as
 $\epsilon \sim 1/[M ( r_s -2M)] $
from Eq. \eqref{red}.
If the observer were dropped at a finite distance but close to the horizon,
for example,
$ |r_s -2M| \ll 1/M$,
the energy density observed by the freely falling observer
would be a huge amount of energy density in this region.
The energy density even in the large black hole can distort the local flatness
in the local inertial frame
due to the contribution of the blue shift to
the energy density compared to the scale of the black hole. This deserves further attention
whether the equivalence principle is still valid or not in this special region.

\section*{Acknowledgments}
We would like to thank Myungseok Eune and Edwin  J. Son for exciting discussions
at the first stage of this work. In particular,
WK would like to thank Sam L. Braunstein, William G. Unruh, and  Joseph Polchinski
for helpful comments, and was greatly indebted to Don Page for many valuable comments
in particular for the explicit explanation of the infinite acceleration at the horizon
in section IV on this work.



\begin{thebibliography}{99}
\bibitem{Hawking:1974sw}
  S.~W.~Hawking,
  Commun.\ Math.\ Phys.\  {\bf 43}, 199 (1975)
  [Erratum-ibid.\  {\bf 46}, 206 (1976)].

\bibitem{Hawking:1976ra}
  S.~W.~Hawking,
  Phys.\ Rev.\  D {\bf 14}, 2460 (1976).

\bibitem{Susskind:1993if}
  L.~Susskind, L.~Thorlacius and J.~Uglum,
  Phys.\ Rev.\  D {\bf 48}, 3743 (1993)
  [arXiv:hep-th/9306069].


\bibitem{Stephens:1993an}
  C.~R.~Stephens, G.~'t Hooft and B.~F.~Whiting,
Class.\ Quant.\ Grav.\  {\bf 11}, 621 (1994)  [gr-qc/9310006].  


\bibitem{Susskind:1993mu}
  L.~Susskind and L.~Thorlacius,
  Phys.\ Rev.\  D {\bf 49}, 966 (1994)
  [arXiv:hep-th/9308100].

\bibitem{Almheiri:2012rt}
  A.~Almheiri, D.~Marolf, J.~Polchinski and J.~Sully,
  JHEP {\bf 1302}, 062 (2013)
  [arXiv:1207.3123 [hep-th]].

\bibitem{Almheiri:2013hfa}
  A.~Almheiri, D.~Marolf, J.~Polchinski, D.~Stanford and J.~Sully,
JHEP {\bf 1309}, 018 (2013)  [arXiv:1304.6483 [hep-th]].

\bibitem{Mathur:2009hf}
  S.~D.~Mathur,
  Class.\ Quant.\ Grav.\  {\bf 26}, 224001 (2009)
  [arXiv:0909.1038 [hep-th]].
%

\bibitem{Bousso:2012as}
  R.~Bousso,
  Phys.\ Rev.\ D {\bf 87}, 124023 (2013)
  [arXiv:1207.5192 [hep-th]].



\bibitem{Nomura:2012sw}
  Y.~Nomura, J.~Varela and S.~J.~Weinberg,
  JHEP {\bf 1303}, 059 (2013)
  [arXiv:1207.6626 [hep-th]].


\bibitem{Page:2012zc}
  D.~N.~Page,
  JCAP {\bf 1304}, 037 (2013)
  [arXiv:1211.6734 [hep-th]].



\bibitem{Giddings:2012gc}
  S.~B.~Giddings,
  Phys.\ Rev.\ D {\bf 88}, 064023 (2013)
  [arXiv:1211.7070 [hep-th]].

\bibitem{JACOBSON:2013ewa}
  T.~Jacobson,
  Int.\ J.\ Mod.\ Phys.\ D {\bf 22}, 1342002 (2013)
  [arXiv:1212.6944 [hep-th]].


\bibitem{Kim:2013fv}
  W.~Kim, B.-H.~Lee and D.-H.~Yeom,
  JHEP {\bf 1305}, 060 (2013)
  [arXiv:1301.5138 [gr-qc]].

\bibitem{Maldacena:2013xja}
  J.~Maldacena and L.~Susskind,
  Fortsch.\ Phys.\  {\bf 61}, 781 (2013)
  [arXiv:1306.0533 [hep-th]].

\bibitem{Hotta:2013clt}
  M.~Hotta, J.~Matsumoto and K.~Funo,
  Phys.\ Rev.\ D {\bf 89}, no. 12, 124023 (2014)
  [arXiv:1306.5057 [quant-ph]].


\bibitem{Braunstein:2013bra}
  S.~L.~Braunstein, S.~Pirandola and K.~Zyczkowski,
  Phys.\ Rev.\ Lett.\  {\bf 110}, 101301 (2013).


\bibitem{Hutchinson:2013kka}
  J.~Hutchinson and D.~Stojkovic,
  arXiv:1307.5861 [hep-th].


\bibitem{Page:2013mqa}
  D.~N.~Page,
  JCAP {\bf 1406}, 051 (2014)
  [arXiv:1306.0562 [hep-th]].

\bibitem{Saini:2015dea}
  A.~Saini and D.~Stojkovic,
  Phys.\ Rev.\ Lett.\  {\bf 114}, no. 11, 111301 (2015)
  [arXiv:1503.01487 [gr-qc]].

\bibitem{Candelas:1980zt}
  P.~Candelas,
Phys.\ Rev.\ D {\bf 21}, 2185 (1980).  




\bibitem{Unruh:1980cg}
  W.~G.~Unruh,
  Phys.\ Rev.\ D {\bf 14}, 870 (1976).

\bibitem{Israel:1976ur}
  W.~Israel,
  Phys.\ Lett.\ A {\bf 57}, 107 (1976).  


\bibitem{Hartle:1976tp}
  J.~B.~Hartle and S.~W.~Hawking,
  Phys.\ Rev.\ D {\bf 13}, 2188 (1976).  

\bibitem{Boulware:1404}
  D.~G.~Boulware,
 Phys.\ Rev.\ D {\bf 11} 1404 (1975).  





\bibitem{Greenwood:2008ht}
  E.~Greenwood and D.~Stojkovic,
  JHEP {\bf 0806}, 042 (2008)
  doi:10.1088/1126-6708/2008/06/042
  [arXiv:0802.4087 [gr-qc]].



\bibitem{Barbado:2011dx}
  L.~C.~Barbado, C.~Barcelo and L.~J.~Garay,
  Class.\ Quant.\ Grav.\  {\bf 28}, 125021 (2011)
  [arXiv:1101.4382 [gr-qc]].



\bibitem{Barbado:2012pt}
  L.~C.~Barbado, C.~Barcelo and L.~J.~Garay,
  Class.\ Quant.\ Grav.\  {\bf 29}, 075013 (2012)
  [arXiv:1201.3820 [gr-qc]].



\bibitem{Eune:2014eka}
  M.~Eune, Y.~Gim and W.~Kim,
  Mod.\ Phys.\ Lett.\ A {\bf 29}, no. 40, 1450215 (2014)
  [arXiv:1401.3501 [hep-th]].





\bibitem{Freivogel:2014dca}
  B.~Freivogel,
  JCAP {\bf 1407}, 041 (2014)
  [arXiv:1401.5340 [hep-th]].

\bibitem{Singh:2014paa}
  S.~Singh and S.~Chakraborty,
  Phys.\ Rev.\ D {\bf 90}, 024011 (2014)
  [arXiv:1404.0684 [gr-qc]].





\bibitem{Ford:1993bw}
  L.~H.~Ford and T.~A.~Roman,
Phys.\ Rev.\ D {\bf 48}, 776 (1993)  [gr-qc/9303038].  



\bibitem{Callan:1992rs}
  C.~G.~Callan, Jr., S.~B.~Giddings, J.~A.~Harvey and A.~Strominger,
  Phys.\ Rev.\ D {\bf 45}, 1005 (1992)
  [hep-th/9111056].











\bibitem{Birrell:1982ix}
 N.D. Birrell, P.C.W. Davies, Quantum Fields In Curved Space, Cambridge Univ.
Press, Cambridge, UK, 1982.






\bibitem{Christensen:1977jc}
  S.~M.~Christensen and S.~A.~Fulling,
   Phys.\ Rev.\ D {\bf 15}, 2088 (1977).  







\end{thebibliography}
\end{document}